\documentclass[prl,tightenlines,preprint,onecolumn]{revtex4}%
\usepackage{amsfonts}
\usepackage{amsmath}
\usepackage{amssymb}
\usepackage{graphicx}
\usepackage{url}%
\begin{document}
\preprint{ }

\begin{center}
{\LARGE Classical Limit of the Casimir Entropy for Scalar Massless Field}

\bigskip

\textit{S. Rubin}$^{a,b}$\textit{, J. Feinberg}$^{b,c},$\textit{ A. Mann}%
$^{b},$\textit{ M. Revzen}$^{b}$

$^{a}$\textit{Department of Physics, Ben-Gurion University of the Negev,
Beer-Sheva 84105, Israel}

$^{b}$\textit{Department of Physics, Technion-Israel Institute of Technology,
Haifa 32000, Israel}

$^{c}$\textit{Department of Physics, Oranim-University of Haifa, Tivon 36006,
Israel}
\end{center}

\bigskip

\begin{center}
{\LARGE Abstract}
\end{center}

We study the Casimir effect at finite temperature for a massless scalar field
in the parallel plates geometry in N spatial dimensions, under various
combinations of Dirichlet and Neumann boundary conditions on the plates. We
show that in all these cases the entropy, in the limit where energy
equipartitioning applies, is a geometrical factor whose sign determines the
sign of the Casimir force.

\bigskip

\bigskip

\bigskip

\bigskip

\bigskip

\bigskip

\bigskip

\bigskip

\bigskip

\bigskip

\bigskip

\bigskip

\bigskip

\bigskip

\bigskip

\bigskip

\bigskip

\bigskip

\bigskip

\bigskip

\bigskip

\bigskip

\bigskip

\bigskip

\bigskip

\bigskip



\bigskip\footnotetext[1]%
{emails:rubinsh@bgu.ac.il,joshua@physics.technion.ac.il,ady@physics.technion.ac.il,revzen@physics.technion.ac.il}%

\section{Introduction}

The vacuum expectation value (VEV) of the Hamiltonian of a free scalar field
in a large volume $V$\ (so that the allowed Fourier modes tend to a continuum)
in $N$\ spatial dimensions at temperature zero is given by the following
expression (given, e.g., in \cite{Zee})%
\begin{equation}%
\begin{array}
[c]{c}%
E_{0}=\left\langle 0\right\vert H_{free}\left\vert 0\right\rangle =\\
\int d^{N}x\int\frac{d^{N}k}{\left(  2\pi\right)  ^{N}2\omega_{k}}\left[
\frac{1}{2}\left(  \omega_{k}^{2}+k^{2}+m^{2}\right)  \right]  =V\int
\frac{d^{N}k}{\left(  2\pi\right)  ^{N}}\frac{1}{2}\omega_{k}\\
\text{where }\omega_{k}^{2}=k^{2}+m^{2},\text{ }k^{2}=%
{\displaystyle\sum\limits_{i=1}^{N}}
k_{i}^{2},
\end{array}
\end{equation}
and we use natural units $\hbar=c=k_{B}=1$. This is the zero point energy of
harmonic oscillators integrated over all momentum modes and over all space.
The sum clearly diverges, but one may define a regularized subtracted
hamiltonian $H^{\prime}=H_{free}-E_{0}$. This simple shift is equivalent to
normal ordering $H^{\prime}=:H_{free}:$. In this way $\left\langle
0\right\vert H^{\prime}\left\vert 0\right\rangle =0$\ for a free field.
Similarly, once the field is constrained by imposing boundary conditions and
thus described by a different hamiltonian $H,$\ its vacuum expectation value
$\tilde{E}_{0}=\left\langle \tilde{0}\right\vert H\left\vert \tilde
{0}\right\rangle $\ diverges ($\left\vert \tilde{0}\right\rangle $\ is the
ground state of the constrained hamiltonian).

It is of physical interest to compute the difference $\tilde{E}_{0}-E_{0}$\ of
the two (divergent) vacuum energies. This was first considered by Casimir (for
the electromagnetic field) \cite{Casimir} and is therefore called the Casimir
energy (and in general the subject is called the Casimir effect). As written,
this difference is not well defined and thus requires regularization. The
properly regularized finite part of that series is, by definition, the Casimir
energy of the constrained system. This is carefully considered in \cite[p
24]{Bordag2} and \cite[p 100]{Pluinen}. We perform the regularization via the
damping function method \cite{Bordag2}. We regularize the energies of the
bounded and of the unbounded system using the same regularizing parameter
$\lambda$. In Eq.(2) we schematically designate this by%
\begin{equation}
E_{Casimir}=\left(  \tilde{E}_{0}-E_{0}\right)  _{reg}. \label{diff cas}%
\end{equation}
At the end of the calculation we take the limit $\lambda=0.$

The Casimir energy depends on the geometry of the constraints. For example, in
Casimir's original work \cite{Casimir} he considered two infinite parallel
plates with separation $d$. In this case, the Casimir energy, $E_{Casimir},$
(per unit area of the plates) is a function of the separation and it gives
rise to a force on the plates, the Casimir force,%
\begin{equation}
F_{Casimir}=-\frac{\partial E_{Casimir}}{\partial d}%
\end{equation}
and this leads to a measurable effect, the Casimir effect.

Various methods may be used to calculate Casimir energies, (e.g., the Green's
function method \cite{Bordag2},\cite{Brown},\cite{Jaffe},\cite{Jaffe2}%
,\cite{Milton},\cite{Mostepanenko},\cite{Pluinen}; path integration \ method
\cite{Feinberg Mann Revzen},\cite{Reuter}; dimensional regularization method
\cite{Elizalde},\cite{Silva}; mode summation method \cite{Casimir}%
,\cite{Milonni},\cite{Milton},\cite{Zuber},\cite{Reuter} etc.). There are
several reviews and books on the subject, e.g., \cite{Bordag2},\cite{Milonni}%
,\cite{Milton},\cite{Mostepanenko},\cite{Pluinen}.

In the present work we consider the Casimir effect for a free massless scalar
field in $N$\ spatial dimensions due to the presence of two parallel
hyperplanes at distance $d,$\ at finite temperature. Hence, instead of the
vacuum expectation values in the definition of the Casimir effect at zero
temperature above, we have to consider expectation values with respect to the
thermal equilibrium density matrix of the system.

We shall consider two cases. In the first case the scalar field is constrained
to vanish on the hyperplanes, imitating the presence of ideal conductors
(Dirichlet-Dirichlet (DD) boundary conditions (b.c.)). In the second case the
field is set to zero on one hyperplane while the normal derivative of the
field is set to zero on the other, thus imitating the presence of a perfect
conductor and a perfectly permeable material (Dirichlet-Neumann (DN) b.c.).

In this paper we mostly apply the mode summation method. In the zero
temperature case we apply the Green's function method as well. Our conclusion
is that the DD case gives rise to attractive forces between the boundaries
while the DN case gives rise to repulsive forces, a result which holds for
both cases for any dimension and any $d>0$\ and $T\geq0$. The case of
Neumann-Neumann (NN) b.c. gives rise to the same effect as the DD case, as we
shall see in the Green's function section. A case of particular interest is to
obtain the force per unit area (pressure) on the boundaries in the high
temperature limit (namely for $T$ such that $\frac{T}{{\large T}_{c}}\gg1$
where $T_{c}=\frac{\hbar c\pi}{k_{B}d})$.

A useful quantity, the Casimir entropy, may be defined \cite{Balian}%
,\cite{Boyer},\cite{Feinberg Mann Revzen},\cite{Revzen2},\cite{Revzen3}. Its
sign in the high temperature limit determines whether the force on the
boundary is attractive or repulsive (or zero if it vanishes). We obtain the
high temperature limit of the Casimir entropy with the following results: For
$N>1,$\ in the DD case the Casimir entropy is negative (attractive force); in
the DN case it is positive (repulsive force). For $N=1$\ we found that the
Casimir entropy is zero in this high temperature limit for both DD and DN b.c.
(the force between the boundary points tends to zero). Calculations for the
Casimir energy for the DN case in 3 spatial dimensions were performed in
\cite{Silva} via zeta function regularization method. The Casimir energy in N
spatial dimensions for the DD case was obtained in \cite{Ambjorn},
\cite{Milton}. We perform such calculation for the DN case in N spatial
dimensions via the mode summation method and discuss the Casimir entropy in
N\ spatial dimensions for both the DD and DN cases.

\section{Mode Summation at Zero Temperature}

The dependence of the sign of the Casimir force on the boundary conditions is
not yet fully understood. It was noted (\cite{Feinberg Mann Revzen},
\cite{Revzen2}) that for simple geometries in the classical limit (i.e., at
temperatures wherein energy equipartitioning is applicable) the total Casimir
force is entropic and the Casimir entropy depends solely on the geometry and
the type of boundary conditions (by the latter we mean various combinations of
Dirichlet and Neumann boundary conditions in the geometry of two parallel
plates). We are interested in calculating the Casimir energy and free energy
of a massless scalar field in the parallel plate geometry of two hyperplanes
located at $x_{N}=0$\ and $x_{N}=d$. Consider $N$\ dimensional space with
coordinate vector $\vec{x}=\left(  x_{1},...,x_{N}\right)  $\ and a cube of
edge length $L$\ in this space with faces given by $x_{i}=-\frac{L}{2}%
,\frac{L}{2}$\ $(i=1,...,N-1)$\ and $x_{N}=-(\frac{L-d}{2}),$\ $\left(
\frac{L+d}{2}\right)  $\ chosen in order to make our volume of interest a
finite one. Later we will take $L\rightarrow\infty,$\ so that only the
boundary conditions (b.c.) on $x_{N}=0,$\ $d$\ will be important (the reason
of treating the coordinate $x_{N}$\ slightly differently is a matter of convenience).

The hyperplanes at $x_{N}=0$\ and $x_{N}=d$\ with b.c. on them divide the cube
into three regions $-(\frac{L-d}{2})<x_{N}<0,$\ $0<x_{N}<d,$\ $d<x_{N}<\left(
\frac{L+d}{2}\right)  .$\ We say that our system is subjected to DD b.c. if on
the hyperplanes $x_{N}=0,d$\ the scalar field is constrained by the conditions
$\left.  \phi\left(  t,\vec{x}\right)  \right\vert _{x_{N}=0}=\left.
\phi\left(  t,\vec{x}\right)  \right\vert _{x_{N}=d}=0.$\ Similarly we say
that our system is subjected to DN b.c. if the scalar field fulfils the
equations $\left.  \phi\left(  t,\vec{x}\right)  \right\vert _{x_{N}%
=0}=\partial_{x_{N}}\left.  \phi\left(  t,\vec{x}\right)  \right\vert
_{x_{N}=d}=0$. For convenience the boundary conditions on the faces
$x_{i}=-\frac{L}{2},$ $\frac{L}{2}$\ $\left(  i=1,...,N-1\right)  $ are taken
to be periodic, which we may write as $\left.  \phi\left(  t,\vec{x}\right)
\right\vert _{x_{i}=-\frac{L}{2}}=$\ $\left.  \phi\left(  t,\vec{x}\right)
\right\vert _{x_{i}=\frac{L}{2}}$\ $\left(  i=1,...,N-1\right)  .$\ As for the
remaining faces of the cube ($x_{N}=-(\frac{L-d}{2}),$\ $\frac{L+d}{2}$) we
impose Dirichlet b.c. $\left.  \phi\left(  t,\vec{x}\right)  \right\vert
_{x_{N}=-(\frac{L-d}{2})}=$\ $\left.  \phi\left(  t,\vec{x}\right)
\right\vert _{x_{N}=\frac{L+d}{2}}=0$\ for simplicity. We expect the details
of the boundary conditions on the faces of the large cube to be unimportant in
the limit $L\rightarrow\infty.$

Following our discussion prior to Eq.(\ref{diff cas})\textbf{ },we define the
Casimir energy (Casimir free energy) of the scalar field as the regularized
difference (in the sense previously discussed) of the total mean energy (free
energy) of the scalar field for the constrained and unconstrained cases.
Introducing the Fourier modes of the field, each mode at temperature $T$
contributes a mean energy%
\begin{equation}%
\begin{array}
[c]{c}%
E_{\vec{k}}\left(  T\right)  =\frac{1}{2}\hbar\omega_{k}\coth(\frac{1}{2}%
\beta\hbar\omega_{k})=\\
\frac{1}{2}\hbar\omega_{k}+\frac{\hbar\omega_{k}}{e^{\beta\hbar\omega_{k}}-1}%
\end{array}
\label{Energy Mode}%
\end{equation}
(where $\beta=\frac{1}{k_{B}T}$) and free energy%
\begin{equation}
F_{\vec{k}}(T)=\frac{1}{2}\hbar\omega_{k}+\frac{1}{\beta}\log\left(
1-\exp\left(  -\beta\hbar\omega_{k}\right)  \right)  .\label{Free Energy mode}%
\end{equation}
At zero temperature both expressions coincide and reduce to
\begin{equation}
E_{\vec{k}}\left(  0\right)  =F_{k}(0)=\frac{1}{2}\hbar\omega_{k}%
\label{Energy mode zero}%
\end{equation}
The angular frequency satisfies $\omega_{k}=ck.$\ The solutions of the free
massless Klein Gordon equation $\square\phi=0$\ in the whole space, for DD and
DN b.c. on the hyperplanes $x_{N}=0$\ and $x_{N}=d,$ give rise to the
following wave vectors in the region $0<x_{N}<d:$
\begin{equation}%
\begin{array}
[c]{c}%
\left\{
\begin{array}
[c]{c}%
\text{DD case: }\vec{k}=\left(  \frac{2\pi}{L}n_{1},...,\frac{2\pi}{L}%
n_{N-1},\frac{\pi}{d}n_{N}\right)  \\
\text{DN case: }\vec{k}=\left(  \frac{2\pi}{L}n_{1},...,\frac{2\pi}{L}%
n_{N-1},\frac{\pi}{d}\left(  n_{N}-\frac{1}{2}\right)  \right)
\end{array}
\right\}  \\
n_{j}=\{0,\pm1,\pm2,....\}\text{ for }j=1,...,N-1\text{ and }n_{N}%
=\{1,2,...\}.
\end{array}
\text{ }\label{k values}%
\end{equation}
Similar expressions hold in the other regions. The total mean energy (free
energy) of a given system is just the regularized sum over the energies (free
energies) of the individual modes. The Casimir energy is defined by
\begin{equation}
{\small E}_{c}\left(  {\small d,T}\right)  {\small =}\left\{  \sum_{\vec
{k}_{1}}{\small E}_{\vec{k}_{1}}\left(  {\small T}\right)  +\sum_{\vec{k}_{2}%
}{\small E}_{\vec{k}_{2}}\left(  {\small T}\right)  +\sum_{\vec{k}_{3}%
}{\small E}_{\vec{k}_{3}}\left(  {\small T}\right)  -\sum_{\vec{k}_{0}%
}{\small E}_{\vec{k}_{0}}\left(  {\small T}\right)  \right\}  _{reg}%
\label{initial EF}%
\end{equation}
and the Casimir free energy is:%
\begin{equation}
{\small F}_{c}\left(  {\small d,T}\right)  {\small =}\left\{  \sum_{\vec
{k}_{1}}{\small F}_{\vec{k}_{1}}\left(  {\small T}\right)  +\sum_{\vec{k}_{2}%
}{\small F}_{\vec{k}_{2}}\left(  {\small T}\right)  +\sum_{\vec{k}_{3}%
}{\small F}_{\vec{k}_{3}}\left(  {\small T}\right)  -\sum_{\vec{k}_{0}%
}{\small F}_{\vec{k}_{0}}\left(  {\small T}\right)  \right\}  _{reg}%
\label{initial Free Energy}%
\end{equation}
where the vectors $\vec{k}_{1},$\ $\vec{k}_{2},$\ $\vec{k}_{3}$\ correspond to
each one of the three regions (the order of regions is from negative $x_{N}%
$\ to positive) of the constrained system, and $\vec{k}_{0}$\ corresponds to
the unconstrained system. The sums in Eqs. (\ref{initial EF}) and
(\ref{initial Free Energy}) are discrete. But because $L$\ is large we can
replace by integrals each of the sums whose summation index is multiplied by
$\frac{2\pi}{L}$\ or $\frac{2\pi}{L-d}$. Those $N-1$ components of $\vec{k}%
$\ become continuous variables of integration and the integrations are
performed over the whole space. In the continuum limit the Casimir energy
(Casimir free energy) in both DD and DN cases is the difference of the total
energy (free energy) of the constrained system in the volume between the
planes and the energy (free energy) of the free system in the volume between
the planes. The damping function we use is the exponential and its explicit
form is given in Eq.(\ref{zero energy}) below. The Casimir energy at zero
temperature due to DD b.c. is given by
\begin{equation}%
\begin{array}
[c]{c}%
E_{c}^{(DD)}\left(  d,0\right)  =\tfrac{L^{N-1}}{2\left(  2\pi\right)  ^{N-1}%
}\times\\
(%
{\displaystyle\sum\limits_{n=1}^{\infty}}
{\displaystyle\int}
{\small d}^{{\small N-1}}k\sqrt{{\small k}^{2}+(\tfrac{\pi n}{d})^{2}}%
\exp(-{\small \lambda}\sqrt{{\small k}^{2}+(\tfrac{\pi n}{d})^{{\small 2}}%
})-\\
\left(  \dfrac{d}{\pi}\right)
{\displaystyle\int\limits_{0}^{\infty}}
{\small dk}_{{\small N}}%
{\displaystyle\int}
d^{N-1}k\sqrt{{\small k}^{2}{\small +k}_{{\small N}}^{2}}\exp(-{\small \lambda
}\sqrt{{\small k}^{2}{\small +k}_{N}^{2}}))
\end{array}
\label{zero energy}%
\end{equation}
where $k^{2}=k_{1}^{2}+...+k_{N-1}^{2}.$ The damping factor $\lambda$ in
Eq.(\ref{zero energy}) serves to regularize the otherwise divergent integrals;
only at the end of the calculation do we take $\lambda$\ to zero. For the
electromagnetic field we can introduce a cutoff damping factor for purely
physical reasons because a physical conductor becomes transparent at high
enough frequencies, namely frequencies higher than the plasma frequency of a
given material. For a scalar field we may regard it as a mathematical device,
since no scalar massless fields were observed in nature so far. The Casimir
energy due to DN b.c. at zero temperature is obtained from (\ref{zero energy})
by shifting the summation index in the integrand $n\rightarrow n-\frac{1}{2}$.
During the calculation $d$-independent terms arise, but we omit them
since\ such $d$-independent terms don't influence physical observables such as
force and pressure \ These terms may be interpreted as self-energy of the
plates, since they are proportional to their area\textbf{ }$L^{N-1}$ . We
denote shifted Casimir energy and shifted Casimir free energy by the same
letters as the unshifted one. The Casimir energy at zero temperature due to DD
b.c. which we obtain is:%
\begin{equation}
E_{c}^{(DD)}\left(  d,0\right)  =-\tfrac{L^{N-1}}{\left(  4\pi\right)
^{\frac{N+1}{2}}d^{N}}\Gamma\left(  \tfrac{N+1}{2}\right)  \zeta\left(
{\small N+1}\right)  \label{Casimir Energy D-D Zero}%
\end{equation}
(which agrees with \cite{Ambjorn},\cite{Elizalde}). On the other hand, for DN
b.c. the Casimir energy is:%
\begin{equation}
E_{c}^{(DN)}\left(  d,0\right)  =\tfrac{L^{N-1}}{\left(  4\pi\right)
^{\frac{N+1}{2}}d^{N}}\Gamma\left(  \tfrac{N+1}{2}\right)  (1-\tfrac{1}{2^{N}%
})\zeta({\small N+1})\label{Casimir energy D-N zero}%
\end{equation}
(which agrees with \cite{Silva}). The Casimir pressure for any b.c. at zero
temperature is given by
\begin{equation}
P_{c}\left(  d,0\right)  =-\tfrac{1}{L^{N-1}}\left(  \dfrac{\partial
E_{c}\left(  d,0\right)  }{\partial d}\right)  .
\end{equation}
We observe that Eq.(\ref{Casimir Energy D-D Zero}) implies attraction between
the hyperplanes, while Eq.(\ref{Casimir energy D-N zero}) implies repulsion.

\section{Mode summation at Finite Temperature}

Let us now turn to the temperature dependent case. We are interested to obtain
the expressions for Casimir energy, Casimir free energy and Casimir entropy in
the cases of DD and DN b.c.. As in the zero temperature case we consider the
differences of energy (free energy) of the constrained system and energy (free
energy) of the free system with mean energy (free energy) per mode given
respectively by Eq.(\ref{Energy Mode}) and Eq.(\ref{Free Energy mode}). We may
decompose the total Casimir energy and total Casimir free energy into a sum of
two terms: the Casimir energy at zero temperature (which we already know) and
a remaining temperature dependent term. In order to obtain expressions for
$E_{c}^{(DD)}\left(  d,T\right)  $ and $F_{c}^{(DD)}\left(  d,T\right)  $ and
$E_{c}^{\left(  DN\right)  }\left(  d,T\right)  ,$ $F_{c}^{\left(  DN\right)
}\left(  d,T\right)  $ we use the Poisson summation formula \cite{Gelfand}%
\begin{equation}
\sum_{n=-\infty}^{\infty}\exp\left(  2\pi inx\right)  =\sum_{n=-\infty
}^{\infty}\delta\left(  x-n\right)
\end{equation}
and our slightly modified version which is obtained by $x\rightarrow
x+\frac{1}{2}$%
\begin{equation}
\sum_{n=-\infty}^{\infty}\left(  -1\right)  ^{n}\exp\left(  2\pi inx\right)
=\sum_{n=-\infty}^{\infty}\delta\left(  x-\left(  n-\frac{1}{2}\right)
\right)  .
\end{equation}
We drop divergent but $d$ independent terms which occur since they don't
change the pressure on the boundaries$.$ After we perform the integrations and
set $\lambda=0$ we replace an infinite series by another infinite series. This
new series expresses the result by known functions (McDonald functions
$K_{\nu}$ \cite{Ryzhik}) and leads to some simplifications of the expressions,
such as cancellation of $E_{c}\left(  d,0\right)  $, and makes easier the
calculation of the high temperature limit ($\beta\rightarrow0$)$.$ The Casimir
energy of a massless scalar field in $N>1$\ dimensions at finite temperature
due to DD and DN b.c. that we obtain is:%
\begin{equation}
E_{c}^{{\small (DD)}}(d,T)=-\tfrac{L^{N-1}\pi}{2^{\frac{N}{2}-4}d^{\frac{N}%
{2}-2}\beta^{\frac{N}{2}+2}}\sum_{m,n=1}^{\infty}\tfrac{n^{\frac{N}{2}+1}%
}{m^{\frac{N}{2}-1}}K_{\frac{N}{2}-1}\left(  \tfrac{4\pi dmn}{\beta}\right)
\label{Energy D-D N dimensions}%
\end{equation}%
\begin{equation}
E_{c}^{{\small (DN)}}(d,T)=-\tfrac{L^{N-1}\pi}{2^{\frac{N}{2}-4}d^{\frac{N}%
{2}-2}\beta^{\frac{N}{2}+2}}\sum_{m,n=1}^{\infty}\left(  -1\right)  ^{m}%
\tfrac{n^{\frac{N}{2}+1}}{m^{\frac{N}{2}-1}}K_{\frac{N}{2}-1}\left(
\tfrac{4\pi dmn}{\beta}\right)  . \label{Energy D-N N dimension}%
\end{equation}
($d$-dependent divergent terms arise during the calculation of the Casimir
energy due to DD and DN b.c. for $N=1$ but they cancel each other at the end.
An independent short calculation based on (\ref{initial EF}) for the $N=1$
case confirms that (\ref{Energy D-D N dimensions}%
),(\ref{Energy D-N N dimension}) are valid for $N=1,$ too.) The double series
converges for all $\beta\geq0$\ and $d>0$, and as $\beta\rightarrow0$\ the
series converges to zero. The conclusion is that the Casimir energy at high
temperature tends to zero in accordance with the following heuristic argument
\cite{Feinberg Mann Revzen}. Each mode carries mean energy given by
Eq.(\ref{Energy Mode})$.$ At the high temperature limit, namely for $T$ such
that $\frac{T}{{\large T}_{c}}\gg1$ where $T_{c}=\frac{\hbar c\pi}{k_{B}d}%
$\ is the geometry dependent temperature scale, we obtain $E_{k}\left(
\beta\rightarrow0\right)  =\frac{1}{\beta}+O\left(  \beta\right)  $, i.e.,
each mode carries the same average energy (which equals the temperature by the
well-known equipartition property of classical physics). The Casimir energy is
the difference of sums of mean energies of the constrained and unconstrained
systems. Moving the hyperplanes $x_{N}=0,$\ $d$\ adiabatically apart will
change the energy levels but at each step there is one to one correspondence
between the energy levels of the systems. Hence zero Casimir energy at the
high temperature limit merely reflects the fact that the number of states of
the constrained and free systems are equal. Generally the Casimir energy due
to equipartition in the high temperature limit is of the form%
\begin{equation}
E_{c}\left(  d,T\rightarrow\infty\right)  =(N_{constrained}-N_{free})\times T
\label{limit N}%
\end{equation}
where $N_{constrained}$ and $N_{free}$ are given by the integrals of the
density of the modes $\rho_{constrained}$ and $\rho_{free}$, and correspond to
the total number of states of the constrained and free systems. In our
infinite parallel planes geometry the mode densities are given in
\cite{Feinberg Mann Revzen} and $N_{constrained}$ and $N_{free}$ turn out to
be both infinite but equal in the sense that the right side of
Eq.(\ref{limit N}) is zero.

The Casimir free energy of a massless scalar field in $N>1$\ spatial
dimensions due to DD and DN b.c is:%
\begin{equation}%
\begin{array}
[c]{c}%
F_{c}^{(DD)}\left(  d,T\right)  =-\frac{\Gamma\left(  \frac{N}{2}\right)
L^{N-1}}{\pi^{\frac{N}{2}}2^{N}\beta d^{N-1}}\zeta\left(  N\right)  -\\
\tfrac{L^{N-1}}{2^{\frac{N}{2}-2}d^{\frac{N}{2}-1}\beta^{\frac{N}{2}+1}}%
{\displaystyle\sum\limits_{m,n=1}^{\infty}}
\frac{n^{\frac{N}{2}}}{m^{\frac{N}{2}}}K_{\frac{N}{2}}\left(  \tfrac{4\pi
mnd}{\beta}\right)
\end{array}
\label{Free Energy D-D N dimensions}%
\end{equation}%
\begin{equation}%
\begin{array}
[c]{c}%
F_{c}^{\left(  DN\right)  }\left(  d,T\right)  =\frac{\Gamma\left(  \frac
{N}{2}\right)  L^{N-1}}{\pi^{\frac{N}{2}}2^{N}\beta d^{N-1}}\left(  1-\frac
{1}{2^{N-1}}\right)  \zeta\left(  N\right)  -\\
\frac{L^{N-1}}{2^{\frac{N}{2}-2}d^{\frac{N}{2}-1}\beta^{\frac{N}{2}+1}}%
{\displaystyle\sum\limits_{m,n=1}^{\infty}}
\left(  -1\right)  ^{m}\frac{n^{\frac{N}{2}}}{m^{\frac{N}{2}}}K_{\frac{N}{2}%
}\left(  \frac{4\pi mnd}{\beta}\right)
\end{array}
\label{Free Energy D-N N dimensions}%
\end{equation}
(the last two formulas give the correct result for the $N=1$\ case, too (see
also below); $\zeta\left(  1\right)  $\ which is divergent is cancelled by an
appropriate term in the double series). The Casimir pressure on the boundary
for any b.c. is given by
\begin{equation}
P_{c}(d,T)=-\tfrac{1}{L^{N-1}}\left(  \dfrac{\partial F_{c}(d,T)}{\partial
d}\right)  _{T}.\label{pressure=dFE/dd}%
\end{equation}
In the DD case it is negative for every temperature and every $d$\ (the
McDonald functions, $K_{\nu},$\ are decreasing functions of their argument).
Hence the Casimir force on the hyperplanes $x_{N}=0,d$\ is attractive while
for the DN case the pressure on the boundary is positive for any $\beta\geq
0$\ and $d>0,$ which implies repulsive force$.$

In the case of one spatial dimension ($N=1$) there are no transverse modes and
the wave vector $\vec{k}$ has only one quantized component. Another way to
obtain the expressions for the $N=1$ case is to promote $N$ to a continuous
variable in (\ref{Energy D-D N dimensions}), (\ref{Energy D-N N dimension}),
(\ref{Free Energy D-D N dimensions}), (\ref{Free Energy D-N N dimensions}),
represent it as $1+\varepsilon$ ($0<\varepsilon\ll1$) and take the limit
$\varepsilon\rightarrow0$. The explicit expressions (which coincide with those
obtained by direct calculation) are:%
\begin{equation}
E_{c}^{\left(  DD\right)  }\left(  d,T\right)  =-\tfrac{\pi d}{\beta^{2}}%
\sum_{n=1}^{\infty}\sinh^{-2}\left(  \tfrac{2d\pi n}{\beta}\right)
\label{EDD N=1}%
\end{equation}%
\begin{equation}
E_{c}^{(DN)}\left(  d,T\right)  =-\tfrac{\pi d}{\beta^{2}}\sum_{n=1}^{\infty
}\left(  -1\right)  ^{n}\sinh^{-2}\left(  \tfrac{2d\pi n}{\beta}\right)
\label{EDN N=1}%
\end{equation}%
\begin{equation}
F_{c}^{\left(  DD\right)  }\left(  d,T\right)  =-\sum_{n=1}^{\infty}\tfrac
{1}{\beta n}\dfrac{1}{\exp\left(  \frac{4\pi nd}{\beta}\right)  -1}
\label{FDD N=1}%
\end{equation}%
\begin{equation}
F_{c}^{\left(  DN\right)  }\left(  d,T\right)  =-\sum_{n=1}^{\infty}\left(
-1\right)  ^{n}\tfrac{1}{\beta n}\dfrac{1}{\exp\left(  \frac{4\pi nd}{\beta
}\right)  -1} \label{FDN N=1}%
\end{equation}

\section{Casimir Entropy and the High Temperature Limit}

Once given the Casimir energy and the Casimir free energy of a system,\ we
define the Casimir entropy \cite{Balian},\cite{Boyer},\cite{Feinberg Mann
Revzen},\cite{Revzen},\cite{Revzen2},\cite{Revzen3} by the equation
\begin{equation}
S_{c}=\dfrac{E_{c}-F_{c}}{T}. \label{Casmir entropy}%
\end{equation}
At the high temperature limit $\left(  \beta\rightarrow0\right)  $ the terms
which contain McDonald functions tend to zero; thus we obtain that the
entropies at the high temperature limit in DD and DN cases for $N>1$ are:%
\begin{equation}
S_{c}^{\left(  DD\right)  }\left(  d,\infty\right)  =\tfrac{\Gamma\left(
\frac{N}{2}\right)  L^{N-1}}{\pi^{\frac{N}{2}}2^{N}d^{N-1}}\zeta\left(
N\right)  =-\beta F_{c}^{\left(  DD\right)  }\left(  d,\infty\right)
\label{SDD}%
\end{equation}%
\begin{equation}%
\begin{array}
[c]{c}%
S_{c}^{\left(  DN\right)  }\left(  d,\infty\right)  =-\frac{\Gamma\left(
\frac{N}{2}\right)  L^{N-1}}{\pi^{\frac{N}{2}}2^{N}d^{N-1}}\left(  1-\frac
{1}{2^{N-1}}\right)  \zeta\left(  N\right)  =\\
-\beta F_{c}^{\left(  DN\right)  }\left(  d,\infty\right)
\end{array}
\label{SDN}%
\end{equation}
(Eq.(\ref{SDD}) coincides with the Casimir free energy in the high temperature
limit given in \cite{Milton}). We see that the Casimir entropy depends on the
geometry (in our case the separation $d$) and on the type of boundary
conditions involved. Its sign is not restricted to be non-negative since the
Casimir entropy is the difference of the entropies of the constrained system
and of the free system and thus might be negative, as happens for the DN b.c..
Recall that in the classical limit the Casimir energy tends to zero due to
equipartition, and therefore in this case the Casimir force depends on the
Casimir entropy only. Hence in the classical limit, the sign of the entropy
determines the sign of the Casimir force. For the two cases of DD and DN b.c.,
we see that the entropies for the two cases differ in sign, thus giving rise
to forces of opposite signs.\textbf{ }Let us observe that $N=1$\ is the only
dimension for which the Casimir entropy at the high temperature limit does not
depend on $d$\ and thus it is an irrelevant constant and the Casimir force is
zero. In the $N=1$ case the boundary consists of two points, which do not
"produce" enough geometry for the Casimir entropy (which gives rise to Casimir
force). Another interesting observation which is deduced from
Eqs.(\ref{Free Energy D-D N dimensions}), (\ref{Free Energy D-N N dimensions})
and (\ref{pressure=dFE/dd}) is that the Casimir pressure $P_{c}(d,T)$
($P_{c}^{\left(  DD\right)  }(d,T)$ or $P_{c}^{\left(  DN\right)  }(d,T)$) has
a fixed sign for fixed $d$ and any $T$ for $N>1$.

\section{Green's Function at Zero Temperature}

Now we turn to the problem of the validity of the mode summation technique.
The correlation function of a massless scalar field at two points separated by
the hyperplane $x_{N}=0,$\ on which the scalar field is subjected to Neumann
boundary conditions ($\left.  \frac{\partial\phi}{\partial x_{N}}\right\vert
_{x_{N}=0}=0$) is not zero but is given by%
\begin{equation}%
\begin{array}
[c]{c}%
\left\langle T\phi\left(  x\right)  \phi\left(  y\right)  \right\rangle
=\Delta_{1}\left(  x-y\right)  +\Delta_{1}\left(  x-\tilde{y}\right)  \\
\text{for }x_{N}>0,\text{ }y_{N}<0
\end{array}
\end{equation}
where $\tilde{y}=(y_{0},y_{1},...,y_{N-1},-y_{N})\ $and $\Delta_{1}\left(
x-y\right)  =\left\langle 0\right\vert \phi\left(  x\right)  \phi\left(
y\right)  \left\vert 0\right\rangle $\ is the homogeneous solution of
$\square_{x}\Delta_{1}\left(  x-y\right)  =0$ (in three spatial dimensions
$\Delta_{1}\left(  x\right)  =-\frac{1}{2\pi^{2}}P.P.\frac{1}{x^{2}}$
\cite{Bjorken}). Since the field operators at points on opposite sides of the
Neumann hyperplane are correlated, it is not possible to expand the field into
independent eigenmodes living on opposite sides of the Neumann hyperplane.
Therefore we should check our assumption of expanding the field into
eigenmodes separately in each one of the regions in the DN case. In contrast,
the correlation function at points on opposite sides of the Dirichlet
hyperplane ($\left.  \phi\right\vert _{x_{N}=0}=0$) is zero. In this case an
expansion of the field into independent eigenmodes living on opposite sides of
the Dirichlet hyperplane is obviously justified. The expression for the
Feynman propagator (which is the time ordered VEV of the two point function
$\left\langle 0\right\vert T\phi\left(  x\right)  \phi\left(  y\right)
\left\vert 0\right\rangle =i\Delta_{F}\left(  x-y\right)  $ in $N$\ spatial
dimensions) is given by \cite{Birrel}%
\begin{equation}%
\begin{array}
[c]{c}%
\Delta_{F}\left(  x,y\right)  _{N,m}=\\
\tfrac{-i\pi}{\left(  4\pi i\right)  ^{\frac{N+1}{2}}}\left(  \tfrac{2m^{2}%
}{-\frac{1}{2}\left(  x-y\right)  ^{2}+i\varepsilon}\right)  ^{\frac{N-1}{4}%
}H_{\frac{N-1}{2}}^{\left(  2\right)  }[\left(  m^{2}\left(  x-y\right)
^{2}-i\varepsilon\right)  ^{\frac{1}{2}}]
\end{array}
\label{Green N dim}%
\end{equation}
where $\left(  x-y\right)  ^{2}=\eta_{\alpha\beta}\left(  x^{\alpha}%
-y^{\alpha}\right)  \left(  x^{\beta}-y^{\beta}\right)  $\ with $\eta
_{\alpha\beta}$\ the diagonal metric of the Minkowski flat space-time with the
following components $\eta_{00}=-\eta_{11}=...=-\eta_{NN}=1$. $H_{\nu
}^{\left(  2\right)  }\left(  x\right)  $\ is the Hankel function
\cite{Arfken}. The leading terms of $\Delta_{F}\left(  x,y\right)  _{N,m}$ as
$m\rightarrow0$ is found to be%
\begin{equation}
\Delta_{F}\left(  x,y\right)  _{N>1,m\rightarrow0}=\frac{2^{N-1}\Gamma\left(
\frac{N-1}{2}\right)  }{i^{N}\left(  4\pi\right)  ^{\frac{N+1}{2}}}\left(
\frac{1}{\left(  x-y\right)  ^{2}}\right)  ^{\frac{N-1}{2}}\label{Green m=0}%
\end{equation}%
\begin{equation}
\Delta_{F}\left(  x,y\right)  _{N=1,m\rightarrow0}=\frac{i}{4\pi}\ln
[m^{2}\left(  x-y\right)  ^{2}]\label{Green m=0, N=1}%
\end{equation}
For $N>1$ it is $m$\ independent and by substituting $N=3$\ into
(\ref{Green m=0}) we obtain the well known result \cite{Bjorken}. For
$N=1$\ we see that as $m\rightarrow0$\ we obtain an infra-red logarithmic
divergent term (infra-red divergence). Actually we may obtain
Eq.(\ref{Green m=0, N=1}) if we promote $N$ to a continuous variable
($N=1+\varepsilon$ ($\varepsilon\ll1$)) in Eq.(\ref{Green m=0}) and keep the
leading terms for small $\varepsilon$ (up to $d$-independent divergent terms)%
\begin{equation}%
\begin{array}
[c]{c}%
\Delta_{F}\left(  x,y\right)  _{N>1,\varepsilon\rightarrow0}=\frac
{\Gamma\left(  \frac{\varepsilon}{2}\right)  }{4\pi i}\left(  \frac{1}{\left(
x-y\right)  ^{2}}\right)  ^{\frac{\varepsilon}{2}}=\\
\frac{\Gamma\left(  \frac{\varepsilon}{2}\right)  }{4\pi i}\exp(-\frac
{\varepsilon}{2}\log\left(  {\small x-y}\right)  ^{2})=\frac{i}{4\pi}%
\log({\small \varepsilon}^{2}\left(  {\small x-y}\right)  ^{2})+O\left(
\varepsilon\right)
\end{array}
\label{e propagator}%
\end{equation}
We see that Eq.(\ref{e propagator}) coincides with Eq.(\ref{Green m=0, N=1})
if we take $\varepsilon=m$.

The free Feynman propagator in $N>1$ spatial dimensions is given by
Eq.(\ref{Green m=0}). Green's functions for DD, DN, NN b.c. in the region
between the planes $0\leq x_{N},$ $y_{N}\leq d$ are given by
Eq.(\ref{Green DD}), (\ref{Green DN}), Eq.(\ref{Green NN}) and in the section
"Optical Green's function" (below) we show how we construct them.
(Eq.(\ref{Green DD}) was previously obtained in \cite{Brown}.)
\begin{equation}%
\begin{array}
[c]{c}%
G^{\left(  DD\right)  }\left(  x,y\right)  =\\%
{\displaystyle\sum\limits_{n=-\infty}^{\infty}}
(\Delta_{F}\left(  x-y+2nd\hat{z}\right)  -\Delta_{F}\left(  x-\tilde
{y}+2nd\hat{z}\right)  )
\end{array}
\label{Green DD}%
\end{equation}%
\begin{equation}%
\begin{array}
[c]{c}%
G^{\left(  DN\right)  }\left(  x,y\right)  =\\%
{\displaystyle\sum\limits_{n=-\infty}^{\infty}}
\left(  -1\right)  ^{n}\left(  \Delta_{F}\left(  x-y+2nd\hat{z}\right)
-\Delta_{F}\left(  x-\tilde{y}+2nd\hat{z}\right)  \right)
\end{array}
\label{Green DN}%
\end{equation}%
\begin{equation}%
\begin{array}
[c]{c}%
G^{\left(  NN\right)  }\left(  x,y\right)  =\\%
{\displaystyle\sum\limits_{n=-\infty}^{\infty}}
\left(  \Delta_{F}\left(  x-y+2nd\hat{z}\right)  +\Delta_{F}\left(
x-\tilde{y}+2nd\hat{z}\right)  \right)
\end{array}
\label{Green NN}%
\end{equation}

Once the Green's function with the proper b.c. is obtained, we may use it to
express the VEV of the energy density of the scalar field in the following way%
\begin{equation}%
\begin{array}
[c]{c}%
\left\langle h\right\rangle =\frac{1}{2}\left\langle \dot{\phi}^{2}\left(
x\right)  +\left(  \vec{\nabla}\phi\left(  x\right)  \right)  ^{2}%
\right\rangle =\\
\frac{1}{2}\lim_{\varepsilon\rightarrow0}[\partial_{x_{0}}\partial_{y_{0}%
}+...+\partial_{x_{N}}\partial_{y_{N}}]\left\langle T\phi\left(  x\right)
\phi\left(  y\right)  \right\rangle
\end{array}
\label{density expectation}%
\end{equation}
where in the last line we regularized the hamiltonian density operator using
the point splitting $x^{\mu}=y^{\mu}+\varepsilon^{\mu}$. The VEV of the total
energy in volume V is:%
\begin{equation}
\left\langle H\right\rangle _{V}=%
{\displaystyle\int\limits_{V}}
\left\langle h\right\rangle dV
\end{equation}
The Casimir energy density is, by definition
\begin{equation}
\left\langle h\right\rangle _{c}=\left\langle h\right\rangle -\left\langle
h\right\rangle _{free}\label{hc}%
\end{equation}
where $\left\langle h\right\rangle $\ is given by
Eq.(\ref{density expectation}) for the constrained field and $\left\langle
h\right\rangle _{free}$\ is the energy density of the free field. The
subtraction in (\ref{hc}) removes the short distance local divergences in
(\ref{density expectation}), which cannot depend on the boundary conditions.
The Casimir energy density is the VEV of the energy density of the field which
is constrained by some boundary conditions, measured relative to the energy
density of a free field. The Casimir energy\ is given by%
\begin{equation}
E_{c}=\left(
{\displaystyle\int\limits_{V}}
\left\langle h\right\rangle _{c}\left(  x\right)  dV\right)  _{reg}+\left(
d\text{ independent}\right)  ,
\end{equation}
where $V$\ is the region between the hyperplanes ($0<x_{N}<d$), and the (maybe
infinite) $d$-independent contribution comes from the infinite region outside
the hyperplanes. Henceforth we shall drop the $d$-independent contribution,
which does not affect the Casimir force. In our three cases the Green's
functions in the region $0\leq x_{N},$\ $y_{N}\leq d$\ are infinite series
(sum over index $n$) which contain the terms $\Delta_{F}\left(  x-y+2nd\hat
{z}\right)  $\ and $\Delta_{F}\left(  x-\tilde{y}+2nd\hat{z}\right)  $\ with
different coefficients, appropriate to the given case. Substitute the
expression for the Green's function into Eq.(\ref{density expectation}). If we
interchange summation over $n$\ and integration with differentiation with
respect to the coordinates we find the contribution of each term to the
Casimir energy. We use the coordinate representation of the free propagator
$\Delta_{F}$\ given by (\ref{Green m=0}), (\ref{Green m=0, N=1}). The
$n=0$\ term which corresponds to the free propagator is cancelled by
Eq.(\ref{hc}). Each "even path" (in the next section "Optical Green's
function", we will explain the terminology "even" and "odd" paths) which is
represented by the term $\Delta_{F}\left(  x-y+2nd\hat{z}\right)  $%
\ ($n\not =0$) contributes to the Casimir energy
\begin{equation}
-\dfrac{\Gamma\left(  \frac{N+1}{2}\right)  }{2\pi^{\frac{N+1}{2}}}\frac
{1}{l_{n}^{N+1}}%
\end{equation}
while each "odd path" which is represented by the term $\Delta_{F}\left(
x-\tilde{y}+2nd\hat{z}\right)  $\ contributes
\begin{equation}
-\dfrac{\Gamma\left(  \frac{N+1}{2}\right)  }{2\pi^{\frac{N+1}{2}}}\frac
{1}{l_{n}^{N+1}}%
\end{equation}
where $l_{n}$ is the length of the path labelled by $n$\ ($n$\ may be any
integer) and is given by
\begin{equation}
l_{n}=\left\{
\begin{array}
[c]{c}%
\left\vert 2dn\right\vert \text{ for "even\ path"}\\
\left\vert 2x_{N}+2dn\right\vert \text{ for "odd path"}%
\end{array}
\right.
\end{equation}
The overall contribution of the "odd" paths to the Casimir energy is
d-independent as one may verify for all three cases of boundary conditions and
therefore unimportant. Consequently, summing over the index $n$\ we obtain the
following well known expressions for Casimir energy (for any $N$)
\cite{Milton}, \cite{Silva}\ which coincide with
Eq.(\ref{Casimir Energy D-D Zero}) and Eq.(\ref{Casimir energy D-N zero}).%
\begin{equation}
E_{c}^{\left(  DD\right)  }\left(  d,0\right)  =-\dfrac{\Gamma\left(
\frac{N+1}{2}\right)  L^{N-1}}{\left(  4\pi\right)  ^{\frac{N+1}{2}}d^{N}%
}\zeta\left(  N+1\right)
\end{equation}%
\begin{equation}
E_{c}^{\left(  DN\right)  }\left(  d,0\right)  =-\left(  1-\dfrac{1}{2^{N}%
}\right)  E_{c}^{\left(  DD\right)  }\left(  d,0\right)
\end{equation}%
\begin{equation}
E_{c}^{\left(  NN\right)  }\left(  d,0\right)  =E_{c}^{\left(  DD\right)
}\left(  d,0\right)
\end{equation}

\section{Optical Green's function}

In this section we are mostly inspired by \cite{Jaffe},\cite{Jaffe2} (which
construct Green's functions for various geometries via classical trajectories
of the rays of light) and \cite{Klichh}. Let us consider a massless scalar
field which is subjected to Dirichlet b.c. on the first hyperplane at
$x_{N}=0$ and Neumann b.c. on the second hyperplane at $x_{N}=d$. Our goal in
this section is to show how one may write the exact Green's function in the
region between the two hyperplanes by making classical, geometric optical
considerations. We shall express the Green's function between the hyperplanes
in terms of propagators connecting two points via paths which hit the
hyperplanes and are reflected (according to the laws of geometric optics). For
each time that the path hits the Dirichlet hyperplane we multiply the
propagator by $(-1)$\ while each time it hits the Neumann hyperplane we
multiply it by $1$. Formally we characterize each path by an index $r=\left(
\sigma_{D}^{(r)},\sigma_{N}^{(r)}\right)  $\ where $\sigma_{D}^{(r)}$\ and
$\sigma_{N}^{(r)}$\ are the reflection numbers, the number of times a given
path $r$\ is reflected from the first and second hyperplane, respectively. The
total number of reflections $\sigma_{D}^{(r)}+\sigma_{N}^{(r)}$\ for path
$r$\ is denoted by $\left\vert r\right\vert .$\ Obviously for every classical
path which we consider, the following relation holds%
\begin{equation}
\left\vert \sigma_{D}^{(r)}-\sigma_{N}^{(r)}\right\vert \leq1.\label{relation}%
\end{equation}
It is convenient to separate the paths into "even paths" and "odd paths".
"Even paths" means paths which are reflected an even number of times from the
hyperplanes. For these paths the sum of reflection numbers, $\left\vert
r\right\vert ,$\ is even and $\sigma_{D}^{(r)}=\sigma_{N}^{(r)}$\ (The last
equality is consistent with (\ref{relation})). "Odd paths" means paths
reflected an odd number of times from the hyperplanes. For odd paths, by
definition, $\left\vert r\right\vert $\ is odd and we may divide them into two
classes according to $\sigma_{D}^{(r)}-\sigma_{N}^{(r)}=\pm1.$\ The total
energy is obtained by summation over all the contributions of all paths. The
total energy due to "even paths" for which $\left\vert r\right\vert =2n$ is:%
\begin{equation}%
\begin{array}
[c]{c}%
E_{even}^{\left(  DN\right)  }=i%
{\displaystyle\int\limits_{V}}
dV\Delta_{F}\left(  x-y\right)  +\\
2i%
{\displaystyle\sum\limits_{n=1}^{\infty}}
\left(  -1\right)  ^{n}%
{\displaystyle\int\limits_{V}}
dV\Delta_{F}\left(  x-y+2nd\hat{z}\right)  =i\sum_{n=-\infty}^{\infty}\left(
-1\right)  ^{n}%
{\displaystyle\int\limits_{V}}
dV\Delta_{F}\left(  x-y+2nd\hat{z}\right)
\end{array}
\label{even contribution}%
\end{equation}
The first factor in Eq.(\ref{even contribution}) corresponds to the path
$\left\vert r\right\vert =0$. The factor 2 is due to degeneracy of the even
paths for which $\left\vert r\right\vert \not =0$. Together they combine to
the r.h.s of Eq.(\ref{even contribution}). The odd paths which first hit the
first hyperplane ( the class $\sigma_{D}^{(r)}-\sigma_{N}^{(r)}=1$) contribute%
\begin{equation}
i\sum_{n=0}^{\infty}\left(  -1\right)  ^{n+1}%
{\displaystyle\int\limits_{V}}
dV\Delta_{F}\left(  x-\tilde{y}+2nd\hat{z}\right)
\end{equation}
while those which hit the second hyperplane first (the class $\sigma_{D}%
^{(r)}-\sigma_{N}^{(r)}=-1$) contribute%
\begin{equation}
i\sum_{n=1}^{\infty}\left(  -1\right)  ^{n+1}%
{\displaystyle\int\limits_{V}}
dV\Delta_{F}\left(  x-\tilde{y}-2nd\hat{z}\right)
\end{equation}
Together they give rise to%
\begin{equation}
E_{odd}^{\left(  DN\right)  }=i\sum_{n=-\infty}^{\infty}\left(  -1\right)
^{n+1}%
{\displaystyle\int\limits_{V}}
dV\Delta_{F}\left(  x-\tilde{y}+2nd\hat{z}\right)
\end{equation}
The contribution to the Casimir energy of the odd paths is actually zero
($E_{odd}^{\left(  DN\right)  }=0$), as we already checked in the previous
section, and only even paths contribute to it. It is given by%
\begin{equation}
E_{c}^{\left(  DN\right)  }=i\sum_{n=-\infty}^{\text{ \ \ \ }\infty\text{
\ \ {\large ,}}}\left(  -1\right)  ^{n}%
{\displaystyle\int\limits_{V}}
dV\Delta_{F}\left(  x-y+2nd\hat{z}\right)  \text{ }\label{DN path}%
\end{equation}
(where the prime means taking the sum without the $n=0$\ term).

One can also easily obtain $E_{c}^{\left(  DD\right)  },$ $E_{c}^{\left(
NN\right)  }$ which coincide with the known results$.$

\section{Summary}

To summarize, we presented in this work calculations for the Casimir energy,
free energy and entropy of a scalar massless field at finite temperature in
the cases of DD and DN boundary conditions. We used the technique of mode
summation for any temperature and also the Green's function method for zero
temperature. In the case of DD boundary conditions we used the usual Poisson
summation formula while in the case of DN boundary conditions we had to modify
it and wrote it in a slightly different form. Later we used an identity for
infinite series to express our results by means of modified Bessel functions
of the second kind (McDonald functions). The new results are the Casimir
energy, Casimir free energy and Casimir entropy due to DN b.c.. At the high
temperature limit (namely when $T\gg T_{c}$) we obtain that Kirchhoff's law
holds, namely, that the Casimir energy tends to zero in both cases, thus it is
independent of the boundary conditions. Another result which we obtain is that
at the high temperature limit the force between the hyperplanes is attractive
for the DD boundary conditions and repulsive in the case of DN boundary
conditions. The fact that the Casimir energy at high temperature limit is zero
but the Casimir force does not vanish indicates that the Casimir force in the
high temperature limit is purely entropic.

The results we obtained in the zero temperature by using Green's functions
coincide with those performed by the mode summation. Further research is
needed to understand why does the mode summation, which inherently assumes
non-correlation between the modes, work in the DN case.

\section{Acknowledgments}

The generous financial support of the Technion is gratefully acknowledged by SR.

\appendix

\end{document}